\documentstyle[aps,preprint,tighten]{revtex}

\newcommand{\lsim}   {\mathrel{\mathop{\kern 0pt \rlap
  {\raise.2ex\hbox{$<$}}}
  \lower.9ex\hbox{\kern-.190em $\sim$}}}
\newcommand{\gsim}   {\mathrel{\mathop{\kern 0pt \rlap
  {\raise.2ex\hbox{$>$}}}
  \lower.9ex\hbox{\kern-.190em $\sim$}}}
\begin{document}

\draft
\preprint{
\begin{tabular}{r}
DFTT 33/95
\\
JHU--TIPAC 95018
\\
hep-ph/9505394
\\
revised version
\end{tabular}
}

\title{Limits on the Neutrino Mass and Mixing Angle \\
{}from Pion and Lepton Decays}

\author{
A. Bottino$^{\mbox{a,b}}$
\footnote{E-mail: bottino@to.infn.it},
N. Fornengo$^{\mbox{c,b}}$
\footnote{E-mail: fornengo@jhup.pha.jhu.edu},
C. W. Kim$^{\mbox{c}}$
\footnote{E-mail: kim@rowland.pha.jhu.edu}
and
G. Mignola$^{\mbox{a,b}}$
\footnote{E-mail: mignola@to.infn.it}
\\ \mbox{ }}

\address{
\begin{tabular}{c}
$^{\mbox{a}}$
Dipartimento di Fisica Teorica, Universit\`a di Torino,
Via P. Giuria 1, 10125 Torino, Italy
\\
$^{\mbox{b}}$
INFN, Sezione di Torino,
Via P. Giuria 1, 10125 Torino, Italy
\\
$^{\mbox{c}}$ Department of Physics and Astronomy,
The Johns Hopkins University,
\\
Baltimore, Maryland 21218, USA.
\end{tabular}
}
\date{June 26, 1995}
\date{revised: December 1, 1995}
\maketitle
\begin{abstract}
Motivated by a recent rather surprising conclusion based on
the 1992 PDG data on the pion, kaon and lepton decays that
if three generations of neutrinos are assumed to be massive
and mixed,
the heaviest neutrino, $\nu_3$,
could have a mass in the range,
$155~\mbox{MeV} \lsim m_3 \lsim 225~\mbox{MeV}$, we have analyzed
the latest 1995 data on the leptonic decays of pion,
$\mu$ and $\tau$ with
the assumption that three generations of neutrinos are massive and
mixed. It is shown that when the radiative corrections are included and
the constraint from partial decay widths is imposed, the 1995 data
are consistent with three massless neutrinos with no mixing. Various
limits on the neutrino mass and mixing angle implied by the 1995
data are presented together with a critique of the previous analysis.
\end{abstract}

\section{Introduction}

In a series of seminal papers \cite{shrock} in the early 1980's,
Shrock proposed a wide range of experimental methods
to obtain possible limits on the neutrino mass and
associated mixing, all based on
the precision analysis of the weak interaction data
on the decays of
the pion, kaon and charged leptons,
$\mu$ and $\tau$. In these papers,
the comprehensive analysis of decay rates
and branching ratios of
the lepton and meson decays  was carried out using the
theoretical framework with three massive
neutrinos and associated
mixing. (It is interesting to note that as early as in 1961
Bahcall and Curtis \cite{bahcall}
proposed a similar method based on pion and muon decays, even before the
discovery of $\nu_\mu$!) At that time, however,
available data were not accurate enough to provide
any significant
results on the limits on the neutrino mass and mixing angle,
in the sense that the limit on the mixing angles were restricted
mostly for large values of the neutrino masses.
In the latter works \cite{yamazaki,PDG94}
the limits were further improved.

The best known
and often quoted
limits on the neutrino mass still come from the
analysis of spectral
shapes in the Kurie plots or other decay kinematics
\cite{books}:

\begin{equation}
\begin{array}{l}
m_1 \equiv m(\nu_1) \lsim 5~\mbox{eV}		\cite{PDG94}	\\
m_2 \equiv m(\nu_2) \lsim 270~\mbox{KeV}	\cite{PDG94}	\\
m_3 \equiv m(\nu_3) \lsim 24~\mbox{MeV} 	\cite{ALEPH}.
\end{array}
\label{kin}
\end{equation}

The results from these
analyses are always
presented with the assumption of  neutrinos with no mixing.
A full analysis of the spectral shapes with three
massive neutrinos with mixing
is very much involved and so far no such analysis
with satisfactory accuracy has been carried out.

In the meantime, much attention has been
focused on entirely different approaches in which
neutrino mass and associated mixing can
be probed indirectly by searching for neutrino
oscillation phenomena. The recent activities
in this approach include the experimental search
for reactor and accelerator neutrino oscillations
and the study of the solar and atmospheric
neutrinos. Although very intriguing indications of
massive neutrinos
with mixing have recently been hinted in the solar
and atmospheric neutrino
experiments and in the LSND experiment,
a definitive answer from these experiments is yet to come.

Recently, Peres, Pleitez and Zukanovich Funchal (PPZ) \cite{PPZ}
carried out a comprehensive analysis
of the existing data on the meson and lepton weak decays
assuming that three generations of neutrinos are massive
and mixed. Their analysis was based on the 1992
Particle Data Group (PDG) data \cite{PDG92} combined with
the latest (in 1993) data on $\tau$ decays \cite{tau:93}. The results
are quite
surprising in that the 1992 data  on the decay branching
ratios would be consistent with a finite mass
for $\nu_3$, i.e.
 155 MeV $ \lsim m_3 \lsim $ 225 MeV. This mass range is
significantly larger than
the most recent upper limit,
$ m_3  \lsim 24 $ MeV
that was obtained from a
kinematical analysis of the $\tau$ decay into five
or six pions and
$\nu_{\tau}$ with no mixing. Moreover, PPZ found
that the mixing angle $\beta$
which represents mixing between $\nu_{1}$ and $\nu_{3}$
is also finite (11$\sim$12 degrees) whereas the mixing
angle $\gamma$
between $\nu_2$ and $\nu_3$ was bounded from above,
thus allowing
zero mixing angle.

Motivated by these rather surprising
results, we have carried out a similar analysis of
the decay rates
and branching ratios of the leptonic decays of the
pion, $\mu$ and $\tau$
with the assumption of massive neutrinos with mixing.
First we have
repeated the PPZ analysis with the same set of the
data (1992 PDG data)
and with the assumptions used by PPZ, confirming
their results. However, we have also found that
their results are significantly modified when the
constraint coming from the partial decay rates,
which PPZ did not use,
is imposed.
The constraint imposed by the decay rate is not an
independent one. Instead, it ensures that possible
fortuitous cancellations in the ratios would not
lead to erroneous conclusions.
Furthermore, we have found that the
radiative corrections,
which PPZ also ignored, are quite important because
the accuracy of the data more than warrants the
inclusion of the radiative corrections in a  precision
analysis such as this.
A similar analysis using the latest 1995 data
shows that the 1995 data is consistent with the picture
of three massless neutrinos
with no mixing.
We have been able to set various limits on the
neutrino masses and mixing angles.

The plan of our paper is as follows. In Section II, we list
the formulas relevant to our analysis. (Some details are
put in Appendix.)  All the data used in our
analysis are collected in Section III, including the data
of 1992 for comparison. In Section IV, we present details on our reexamination
and critique of the PPZ analysis. In particular,
we discuss here what would
happen to the PPZ
conclusions based on the  1992 PDG data,
when the radiative corrections are included and the constraint
{}from the partial
decay rates is imposed. New results  based on the 1995 data are presented
in Section V, and a summary and conclusions are given in Section VI.

\section{Formulas with Three Generation Mixing}

The mixing matrix $V$ in the lepton sector
which relates the (weak) interaction eigenfields
$\nu_\alpha$ $(\alpha=e, \mu, \tau)$ to the mass eigenfields
$\nu_i$ $(i=1,2,3)$ is given by

\begin{equation}
\nu_\alpha = \sum_{i=1}^3 V_{\alpha i} \nu_i
\end{equation}

We parameterize the mixing matrix $V$ using the Maiani
representation \cite{maiani}
of the mixing matrix $U$ in the quark sector, with
CP--violating phase set to zero, i.e.

\begin{equation}
V = \left(
\begin{array}{ccccccc}
c_\theta c_\beta & \mbox{ } & \mbox{ } &
 s_\theta c_\beta & \mbox{ } & \mbox{ } &  s_\beta       \\
 & & & & & & \\
 -s_\theta c_\gamma -c_\theta s_\gamma s_\beta & \mbox{ } & \mbox{ } &
 c_\theta c_\gamma -s_\theta s_\gamma s_\beta & \mbox{ } & \mbox{ } &
s_\gamma c_\beta \\
  & & & & & &\\
 s_\theta s_\gamma -c_\theta c_\gamma s_\beta
 & \mbox{ } & \mbox{ }
 & -c_\theta s_\gamma -s_\theta c_\gamma s_\beta
& \mbox{ } & \mbox{ } &c_\gamma c_\beta
\end{array}
\right).
\label{matrix}
\end{equation}
where $s_\theta\equiv \sin\theta$, $c_\theta\equiv \cos\theta$,
$s_\beta\equiv \sin\beta$~\dots. In Eq.(\ref{matrix}),
the angle $\theta$ refers to  mixing between
$\nu_1$ and $\nu_2$, $\beta$ to $\nu_1$ and $\nu_3$ and
$\gamma$ to $\nu_2$ and $\nu_3$, respectively.

\bigskip

In our analysis, we will discuss the limits on neutrino
masses and mixing angles which can be inferred from
purely leptonic decays of pion and leptons ($\mu $
and $\tau$). That is, the decay rates to be used are

\begin{eqnarray*}
\begin{array}{c}
\Gamma(\pi \longrightarrow e \bar\nu_e),~~
\Gamma(\pi \longrightarrow \mu \bar\nu_\mu),
\\
\Gamma(\mu \longrightarrow e \bar\nu_e \nu_\mu),
\\
\Gamma(\tau \longrightarrow e \bar\nu_e \nu_\tau),~~
\Gamma(\tau \longrightarrow \mu \bar\nu_\mu \nu_\tau).
\end{array}
\end{eqnarray*}

These are the best known experimental quantities
which do not involve hadrons in the
final states, hence introducing
 no further unnecessary
complications in the calculation
of decay widths. We will not consider the $K$ decays, even
though some experimental determination of its decay widths into
leptons are almost as good as
those of the pion. Its properties are quite
similar to those of the pion and its data
do not provide any additional
(or critical) information.

Here, we briefly summarize the formulas to be used. (Details
are given in Appendix.)
For the pion, the decay rate into two leptons
in a general case of three massive
neutrinos with mixing is given by

\begin{equation}
\Gamma(\pi \longrightarrow l \bar\nu_l) =
\frac{G^2 f_\pi^2 U_{ud}^2 m_\pi^3}{8 \pi}
\cdot {\cal R}_{\pi l} \cdot
\sum_{i=1}^3 |V_{li}|^2 P_i^{\pi l},
\label{pion:gamma}
\end{equation}

\noindent
where $G$ is the Fermi constant (see comment below),
$m_\pi$ is the pion mass,
$f_\pi$ is the pion decay constant and
$U_{ud}$ is the $ud$ component of the mixing matrix
in the quark sector. The matrix element--phase space function,
$P_i^{\pi l}$, denotes the quantity of our interest which
contains part of the matrix element and the entire phase space,
and it depends on the neutrino masses, $m_{i}$ ($i =1,2,3$),
as well as on the pion and lepton masses. ${\cal R}_{\pi l}$
is a factor that represents the radiative corrections to the process.
We stress
here that ${\cal R}_{\pi l}$ depends on the pion mass
as well as on the lepton mass.
The complete expressions for $P_i^{\pi l}$ and
${\cal R}_{\pi l}$ are given in Appendix.
We wish to emphasize here that the value of $f_\pi = (130.7 \pm
0.1 \pm 0.36)~\mbox{MeV}$ quoted in PDG data set is obtained from the
decay $\pi \longrightarrow \mu \nu_\mu + \mu \nu_\mu \gamma$
under the hypothesis that the
neutrinos are massless with no mixing. For massive and
mixed neutrinos, the above value represents the quantity
$\big [f_\pi^2 \sum_{i=1}^3 |V_{\mu i}|^2 P_i^{\pi \mu} / P_0^{\pi \mu}
\big ]^{1/2}$ rather than $f_\pi$, where $P_0^{\pi \mu}$ is the
matrix element--phase space function for massless neutrinos.
For this reason, in the following, we will consider,
as was done by PPZ, only the ratio of the two leptonic
decay widths of the pion,
in order to cancel out the dependence on the unknown quantity $f_\pi$.

The decay width for a lepton decaying into three leptons is
given by
\begin{equation}
\Gamma(l' \longrightarrow l \bar \nu_{l} \nu_{l'}) =
\frac{G^2 m_{l'}^5}{192 \pi^3}\cdot {\cal R}_{l'} \cdot
\sum_{i,j=1}^3 |V_{l'i}|^2 |V_{lj}|^2 P_{ij}^{l'l},
\label{gamma:l}
\end{equation}

\noindent
where $m_{l'}$ is the mass of the decaying particle.
Again, $P_{ij}^{l'l}$ is the matrix element--phase space
function which depends on the masses of all the particles
involved in the decay process. The leading radiative corrections
are denoted by ${\cal R}_{l'}$; they depend only on the mass
of the decaying lepton. Also, the expressions
for $P_{ij}^{l'l}$ and ${\cal R}_{l'}$ are collected in
Appendix.

Again, it is crucial to emphasize, as noted by PPZ,
that the experimental value of
$G_\mu$ quoted in the PDG data set becomes the Fermi constant
$G$ only for the massless neutrinos, as implied by the Standard
Model. Since new physics beyond the Standard Model is what we
wish to investigate, the coupling constant $G_\mu$ that one measures
in the muon decay should be interpreted as

\begin{equation}
\begin{array}{rl}
G^2_\mu =
& \left [\frac{\mbox{$G^2$}}{\mbox{$P^{\mu e}_{00}$}}\right ] \times
\displaystyle \sum_{i,j=1}^3 |V_{\mu i}|^2 |V_{e j}|^2 P^{\mu e}_{ij}
\\
& \\
= & \left [\frac{\mbox{$G^2$}}{\mbox{$P^{\mu e}_{00}$}}\right] \times
\Big \{
c_\theta^2 c_\beta^2 (s_\theta c_\gamma + c_\theta s_\gamma s_\beta)^2
P^{\mu e}_{11}
\\
 & +
c_\beta^2 [ c_\theta^2 (c_\theta c_\gamma + s_\theta s_\gamma s_\beta)^2 +
s_\theta^2 (s_\theta c_\gamma + c_\theta s_\gamma s_\beta)^2 ]
P^{\mu e}_{12}
\\
 & +
[ c_\beta^2 (c_\theta^2 c_\beta^2 s_\gamma^2) +
s_\beta^2 (s_\theta c_\gamma + c_\theta s_\gamma s_\beta)^2 ]
P^{\mu e}_{13}
\\
 & +
s_\theta^2 c_\beta^2 (c_\theta c_\gamma + s_\theta s_\gamma s_\beta)^2
P^{\mu e}_{22}
\\
 & +
[ c_\beta^2 (s_\theta^2 c_\beta^2 s_\gamma^2) +
s_\beta^2 (c_\theta c_\gamma + s_\theta s_\gamma s_\beta)^2 ]
P^{\mu e}_{23}
\\
 & +
s_\beta^2 c_\beta^2 s_\gamma^2
P^{\mu e}_{33}
\Big \}
\\
& \\
& \longrightarrow
G^2~~~~\mbox{as $m_1, m_2, m_3 \longrightarrow 0$}
\end{array}
\label{Gmu:G}
\end{equation}

\noindent
where $P^{\mu e}_{00}$ is the matrix element--phase space function
for massless neutrinos.
Therefore, in a general case of massive neutrinos with mixing, the weak
coupling constant $G$, which enters in the calculation of all the weak
processes like e.g. Eq(\ref{pion:gamma})
and Eq.(\ref{gamma:l}), is \underline{\em not} directly
measured from the muon decay. The quoted number
$G_\mu = (1.16639 \pm 0.00002) \cdot 10^{-5}~\mbox{GeV}^{-2}$ is valid
only for massless neutrinos.

An alternative way to obtain $G$ comes from the
Standard Model of electroweak interactions, where $G$,
the fine--structure constant $\alpha$, the weak mixing angle
$\theta_W$ and the $W$ boson mass are related as \cite{PDG94,onshell}

\begin{equation}
G = \frac{\pi \alpha}{\sqrt{2} m_W^2 \sin^2\theta_W (1 - \Delta r)}.
\label{G:SM}
\end{equation}

The above equation relates the low-energy quantities $G$ and $\alpha$ to the
quantities defined at the electroweak scale, $\sin^2\theta_W$ and $m_W$.
The radiative effects are taken into account in $\Delta r$, which in
the first order in $\alpha$ can be cast in the following way

\begin{equation}
\Delta r = \Delta r_\alpha + \Delta r_t + \Delta r_f + \Delta r_q +
\Delta r_b
\label{deltar}
\end{equation}

The first term in Eq.(\ref{deltar}) describes the running of $\alpha$
{}from the zero--momentum scale to the electroweak scale, the second
term is due to the radiative contribution of the top, $\Delta r_f$ and
$\Delta r_q$ are contributions coming from the leptons and quarks (except
the top) and the last one is due to the Higgs--boson loops.
The actual form of the radiative correction $\Delta r$ depends on the
renormalization scheme adopted to perform the calculations.

Hence, in the Standard Model only three out of the four quantities
$G$, $\alpha$, $\sin^2\theta_W$ and $m_W$ are independent. Since the
Fermi constant (for massless neutrinos), the fine--structure constant
and the Weinberg angle are the best known quantities, the standard
practice is to consider them as being independent, and then
use Eq.(\ref{G:SM}) to predict the W mass.

The aim of our analysis is to investigate the possibility that neutrinos
are massive and mixed, therefore $G$, as was discussed, is different
{}from $G_\mu$ and hence should be treated as unknown.
Namely, we reverse the standard procedure and use Eq.(\ref{G:SM})
to calculate the value and the allowed 1--$\sigma$ range for the Fermi
constant, using $\alpha$, $\sin^2\theta_W$ and $m_W$ as input parameters.
Obviously we expect the accuracy of the value $G$ to be
rather poor, i.e. of the same order of the experimental error on $m_W$. It
is encouraging, however, that the latest measurement on the $W$ mass are
at the level of 2\% \cite{mw}.

Special caution must be exercised in choosing the value of $\sin^2\theta_W$
to be used in Eq.(\ref{G:SM}). This parameter has been obtained to a
very good accuracy by a number of different experiments (see,
e.g. Ref. \cite{PDG94} and references quoted therein,
for a summary of the different techniques used to extract $\sin^2\theta_W$).
All these methods involve the measurement of cross sections, and the
fits of the data (from which $\sin^2\theta_W$ is obtained) are made using
$G$ for massless neutrinos ($G=G_\mu$) as an input parameter
(directly in the cross section evaluations,
or indirectly through the interference terms in the calculations
of the asymmetries). This means that the best mean value of
$\sin^2\theta_W$ quoted in the PDG data cannot be used in Eq.(\ref{G:SM}),
because it is not independent of $G$ itself. Therefore, we
perform our calculation of $G$ in the on--shell scheme of renormalization
\cite{onshell},
where the following relation is defined to be true at all orders in the
perturbation theory

\begin{equation}
\sin^2\theta_W = 1 - \frac{m_W^2}{m_Z^2}.
\label{SW:onshell}
\end{equation}

In this way $\sin^2\theta_W$ is obtained independently of $G$, even though
its uncertainties will become larger than that of the best mean value
in PDG data due to the uncertainties in the weak boson masses.
(A further comment: strictly speaking, the determination of
$m_W$ and $m_Z$ could, in principle, be also affected
by the value of $G$. In fact, the usual way to determine the weak boson
masses involves a global fit of a number of physical quantities
and $G$ is again taken as an input
parameter of the fit \cite{yellow}.
Therefore, the actual value of $G$ could influence
very slightly the determination of $m_W$ and $m_Z$. However,
the correlation between the value of $G$ and the position of the poles in
the $W$-- and $Z$--production cross--sections (basically the weak boson
masses) cannot be substantial. For this reason we assume the quoted measured
values of $m_W$ and $m_Z$ as being independent of $G$.)

In summary, we have chosen to determine the value of the Fermi constant and
its 1--$\sigma$ allowed interval by means of the Standard Model relation
Eq.(\ref{G:SM}), in the framework of the on--shell renormalization scheme
where $\sin^2\theta_W$ is defined by Eq.(\ref{SW:onshell}). The actual
expressions for the radiative corrections $\Delta r$ in the on--shell
scheme can be found in Ref.\cite{onshell}. The top contribution
$\Delta r_t$, not contained in the original paper by Sirlin, is
\cite{PDG94}

\begin{equation}
\Delta r_t = - \frac{3 \alpha}{16 \pi \sin^2\theta_W}
\frac{m_t^2}{m_W^2} \frac{1}{\tan^2\theta_W}.
\end{equation}

Hence our input parameters are: $m_Z$, $m_W$ and $\alpha$. (Obviously,
the low--energy value of the fine structure constant, obtained from the
quantum Hall effect, is independent from $G$). Some additional input
parameters, which enter in the radiative corrections $\Delta r$, are
the mass of the top quark $m_t$ and the Higgs boson mass $m_H$.
For the top mass, we use the recent CDF
measurement $m_t = (176 \pm 18)~\mbox{GeV}$
\cite{top}, both for the 1992
and 1995 data sets. The Higgs mass is varied in the interval
(60~\mbox{GeV}, 1~\mbox{TeV}).

In the case of the 1992 data set, we use the 1992 PDG values for
$m_Z$ and $m_W$. The results are (here and hereafter, the errors
are propagated quadratically)

\begin{equation}
\begin{array}{l}
\sin^2\theta_W = 0.2258 \pm 0.0050      \\
\Delta r = 0.0402 \pm 0.0088            \\
G = (1.162 \pm 0.029) \cdot 10^{-5}~\mbox{GeV}^{-2}
\end{array}
\end{equation}

For the 1995 data, in order to reduce as much as possible the
uncertainties in the determination of $G$, we use the latest
data available:
$m_Z = (91.1884 \pm 0.0022)~\mbox{GeV}$ \cite{mz} and
$m_W = (80.410 \pm 0.180)~\mbox{GeV}$ \cite{mw}. This
gives

\begin{equation}
\begin{array}{l}
\sin^2\theta_W = 0.2224 \pm 0.0035      \\
\Delta r = 0.0396 \pm 0.0088            \\
G = (1.174 \pm 0.022) \cdot 10^{-5}~\mbox{GeV}^{-2}
\end{array}
\end{equation}

The values of $G$ which we will use in the evaluation of the leptonic decay
widths are those given in the two previous equations.
Unfortunately,
the errors on $G$ (of the order of percent) are much worse than the errors
on $G_\mu$ which are of the order of $10^{-5}$. We stress that since we
are interested in the investigation of massive and mixed neutrinos, the
method discussed above turns out to be the only viable way to determine the
Fermi constant. Unfortunately, however, this procedure determines $G$
with errors of a few percent level and thus makes the accuracy in
the calculation of the
decay widths rather poor. Nevertheless, as will be seen in the following,
it is still possible to use them as an important constraint on the neutrino
parameters (mainly on the neutrino masses). In fact,
a deviation (if any) of $G_\mu$ from the Standard
Model value of $G$ given in Eq.(\ref{G:SM}) signals
the massive and mixed neutrinos.

Measured experimentally are the branching ratios
of the above decay processes.
The branching ratios are simply related to the previously defined
quantities as

\begin{equation}
\mbox{BR}(\pi \longrightarrow l \bar\nu_l) =
\frac{\Gamma(\pi \longrightarrow l \bar\nu_l)}{\Gamma_\pi}
= \tau_\pi \cdot \Gamma(\pi \longrightarrow l \bar\nu_l)
\end{equation}

\begin{equation}
\mbox{BR}(l' \longrightarrow l \bar\nu_l \nu_{l'}) =
\frac{\Gamma(l' \longrightarrow l \bar\nu_l \nu_{l'})}{\Gamma_{l'}}
= \tau_{l'} \cdot \Gamma(l' \longrightarrow l \bar\nu_l \nu_{l'})
\end{equation}

\noindent
where $\Gamma_\pi$ and $\Gamma_{l'}$ are the total widths of
the pion and the decaying lepton, respectively, and $\tau_\pi$
and $\tau_{l'}$ are the corresponding lifetimes.
In order to directly extract information about the neutrino
mass and mixing, instead of the lepton decay widths themselves, we use
the following quantities which are simply proportional to the decay
widths, with common constants such as $G$ and $m_{l'}$ removed,

\begin{equation}
\overline{\Gamma}^{l'l} = \alpha_{l'}
\mbox{BR}(l'\longrightarrow\l\bar\nu_l \nu_{l'}) =
{\cal R}_{l'} \cdot
\sum_{i,j=1}^3 |V_{l'i}|^2 |V_{lj}|^2 P_{ij}^{l'l},
\end{equation}
where

\begin{equation}
\alpha_{l'} = \frac{192\pi^3}{G^2 m_{l'}^5 \tau_{l'}}.
\label{alphal}
\end{equation}

\noindent
{}From the experimental values of the $\mbox{BR}$'s and the physical
quantities defined in Eqs. (\ref{alphal}),
we will first calculate the 1--$\sigma$ allowed experimental ranges
for the $\overline{\Gamma}$'s. These ranges will then be compared with
the calculated values of the $\overline{\Gamma}$'s by varying
the neutrino masses and mixing angles. This would limit the neutrino
masses and mixing angles.

We will follow the same procedure for some ratios of
the $\overline{\Gamma}$'s and the BR's. Our choice of the ratios is

\begin{equation}
R^{\pi e}_{\pi\mu} \equiv
\frac{\mbox{BR}(\pi \longrightarrow e \bar\nu_e)}
{\mbox{BR}(\pi \longrightarrow \mu \bar\nu_\mu)} =
\frac{{\cal R}_{\pi e} \cdot \sum_i |V_{ei}|^2 P_i^{\pi e}}
{{\cal R}_{\pi \mu} \cdot \sum_i |V_{\mu i}|^2 P_i^{\pi \mu}},
\end{equation}

\begin{equation}
R^{l_1'l_1}_{l_2'l_2} \equiv
\frac{\overline\Gamma^{l'_1 l_1}}{\overline\Gamma^{l'_2 l_2}}
=
\left(
\frac{m_{l'_2}^5 \tau_{l'_2}}{m_{l'_1}^5 \tau_{l'_1}}
\right)
\frac
{\mbox{BR}(l'_1 \longrightarrow l_1 \bar\nu_{l_1} \nu_{l'_1})}
{\mbox{BR}(l'_2 \longrightarrow l_2 \bar\nu_{l_2} \nu_{l'_2})} =
\frac{{\cal R}_{l_1'} \cdot
 \sum_{i,j} |V_{l_1'i}|^2 |V_{l_1j}|^2 P_{ij}^{l_1'l_1}}
{{\cal R}_{l_2'} \cdot
 \sum_{i,j} |V_{l_2'i}|^2 |V_{l_2j}|^2 P_{ij}^{l_2'l_2}}.
\end{equation}

In the case of lepton decays,
we will perform our analysis using the two ratios, $R^{\mu e}_{\tau e}$
and $R^{\tau e}_{\tau \mu}$. The use of ratios alone is indeed simpler
because uncertainties in some constant quantities are cancelled out, but
some changes in the numerator and the denominator coming from phase
space and mixing angles may partially be compensated. Therefore one
must check that the calculated single partial decay widths do not
lie outside the experimental allowed ranges. This is why we will add
as an additional constraint also the three leptonic decay widths.
Summarizing, we will use the following quantities as constraint:
$\overline{\Gamma}^{\mu e}$, $\overline{\Gamma}^{\tau e}$,
$\overline{\Gamma}^{\tau \mu}$, $R^{\pi e}_{\pi \mu}$,
$R^{\mu e}_{\tau e}$ and $R^{\tau e}_{\tau \mu}$

When we evaluate the $\overline{\Gamma}$'s and the $R$'s from the experimental
values, we propagate the errors quadratically. In the
calculations, we use the central values
of masses of the particles involved.
All the calculations are carried out at 1--$\sigma$ level.
Also, it is to be noted that  the radiative correction
factors ${\cal R}$'s are included in the quantities
$\overline{\Gamma}$'s; they are cancelled out
in   $R^{\tau e}_{\tau \mu}$ but not in
$R^{\mu e}_{\tau e}$ and $R^{\pi e}_{\pi \mu}$.

\section{Numerical Inputs}

We have listed, in Table 1, all the latest (1995 data)
experimental inputs that will be used in our analysis \cite{PDG94,tau}.
In order to compare our new analysis with the previous one by PPZ
based on the 1992 PDG data,
we have also listed the 1992 PDG data \cite{PDG92}
in Table 1. As can be seen in Table 1, the entries with asterisks
signify those with noticeable changes from the 1992 data to the 1995 data.
In particular, they include
all the data on $\tau$ decays and $\pi \rightarrow e\overline
{\nu}_e$. For the sake of comparison, in Table 1
the value of both $G_\mu$ and $G$ are included.
(We note that, strictly speaking, $m_\tau$ quoted in Table 1 is
for massless $\nu_\tau$. However, in this case, the use of $m_\tau$
in Table 1 does not introduce any significant modification.)

We should notice that the errors on the $\overline{\Gamma}$'s
are quite large (of the order of a few percent) due to the
uncertainties in the Fermi constant $G$, as discussed in the
previous Section. Nevertheless, the inclusion of the
$\overline{\Gamma}$ constraint turns out to be effective in limiting
the allowed intervals for the neutrino parameters, in particular the
masses, as will be shown in the following Sections.

Listed below are the $calculated$ values of $\overline{\Gamma}$'s and
$R$'s using the latest mass values for the decaying
particles and charged leptons under the assumption that
neutrinos are massless with no mixing.
In the case of $\overline{\Gamma}$'s, the first numbers on the right--hand
side are the values without the radiative corrections and
the second numbers represent the radiative corrections.
All the values are in agreement with the 1995 data in Table 1
within 1--$\sigma$,
implying that the 1995 data are consistent with the lepton sector with
massless neutrinos with no mixing.

\bigskip
\bigskip
\centerline{\bf Calculated values for massless neutrinos with no mixing}
\centerline{\bf (1995 data set)}
\bigskip
\bigskip

$
\begin{array}{lll}
 \overline{\Gamma}^{\mu e} = 0.999813 \cdot 0.995797 = 0.995611
& \mbox{~~~~~} &
 R^{\pi e}_{\pi \mu} = 1.233 \cdot 10^{-4}
\\
 \overline{\Gamma}^{\tau e} = 1.00 \cdot 0.996 = 0.996
& \mbox{~~~~~} &
 R^{\mu e}_{\tau e} = 0.9995
\\
 \overline{\Gamma}^{\tau \mu} = 0.972 \cdot 0.996 = 0.968
& \mbox{~~~~~} &
 R^{\tau e}_{\tau \mu} = 1.028
\end{array}
$

\bigskip

(The uncertainties in the above numbers due to the experimental
errors of the quantities which enter in their calculations are
always less than $0.01\%$).

The conclusion based on the above numbers that the experimental
data (1995) are consistent with the assumption of massless neutrinos
with no mixing is in sharp contrast to the result coming from
the 1992 PDG data, as obtained by PPZ.
In order to further examine 1992 data set, we have repeated the
calculations using the 1992 PDG data both with and without the
inclusion of the radiative corrections. We list, in the following,
the results of the $calculated$ values based on the 1992 PDG data
in Table 1 with
the assumption that neutrinos are massless  with no mixing.

\bigskip
\bigskip
\centerline{\bf Calculated values for massless neutrinos with no mixing}
\centerline{\bf (1992 PDG data set)}
\bigskip
\bigskip

$
\begin{array}{l}
\overline{\Gamma}^{\mu e}    =  0.999813\cdot 0.995797  =  0.995611  \\
\overline{\Gamma}^{\tau e}   = 1.00 \cdot 0.996 =  0.996            \\
\overline{\Gamma}^{\tau \mu} =  0.973 \cdot 0.996  =  0.969
\end{array}
$

\bigskip

$
\begin{array}{llll}
R^{\pi e}_{\pi \mu} = 1.233 \cdot 10^{-4}
& \mbox{~~~} &
(R^{\pi e}_{\pi \mu} = 1.283 \cdot 10^{-4}
& ~~~\mbox{without RC})
\\
R^{\mu e}_{\tau e} = 0.9995
& \mbox{~~~} &
(R^{\mu e}_{\tau e} = 0.9998
& ~~~\mbox{without RC})
\\
R^{\tau e}_{\tau \mu} = 1.028
& \mbox{~~~} &
(R^{\tau e}_{\tau \mu} = 1.028
& ~~~\mbox{without RC})
\end{array}
$

\bigskip

(Again, the uncertainties in the above numbers are also less than
$0.01\%$).

Note that  the  calculated values of the following
quantities do $not$ lie in their corresponding
experimental 1--$\sigma$ ranges:
$R^{\pi e}_{\pi \mu}$ and $R^{\mu e}_{\tau e}$.
This implies that the 1992 PDG data are indeed incompatible,
at least within 1--$\sigma$,  with
the assumption that neutrinos are massless with no mixing.
If the radiative corrections are not included,
$\overline{\Gamma}^{\tau e}$ is not compatible, also.

\section{Reexamination of Previous Analysis}

In this Section we will first reexamine in detail the result of the
PPZ analysis and then present the results of our new analysis.
As mentioned already, the PPZ analysis was based on the 1992 PDG data,
improved by the  latest (at that time) determination of the $\tau$
mass. In order to cancel out the dependence of the
decay widths on some parameters (the pion mass, the quark
mixing angle $U_{ud}$, the pion decay constant, the Fermi
constant and the muon or tau mass), PPZ considered only the ratios of the
partial decay widths; $R^{\pi e}_{\pi\mu}$, $R^{\mu e}_{\tau e}$
and $R^{\tau e}_{\tau\mu}$. They did not take
into account radiative
corrections for these processes under the
assumption that the radiative corrections of order of several
percents are of no importance (note that
the radiative corrections do not
cancel each other in the ratios $R^{\pi e}_{\pi\mu}$ and
$R^{\mu e}_{\tau e}$). The PPZ analysis was performed in the
case of one massive neutrino ($\nu_3$) and two almost degenerate
very light ($m_1 \sim m_2 \ll m_3$) ones. Their main result is that $\nu_3$
could have a mass in the interval

\begin{equation}
155~\mbox{MeV} \lsim m_3 \lsim 800~\mbox{MeV}.
\label{PPZlim}
\end{equation}
That is, all the 1992 PDG data could be fitted with $m_3$ in the above
range. They then improved
the upper limit in Eq.(\ref{PPZlim}) by  taking into account the constraint
coming from the $Z$ invisible width. The resulting allowed interval
was

\begin{equation}
155~\mbox{MeV} \lsim m_3 \lsim 225~\mbox{MeV}.
\label{mass}
\end{equation}

Given the values of $m_3$ inside this range, the PPZ analysis
also showed that one
of the mixing angles (namely $\beta$) was constrained
to a finite range which did not include $\beta=0$. They obtained,
for $m_3=165~\mbox{GeV}$,
\begin{equation}
11.54^\circ \lsim \beta \lsim 12.82^\circ.
\label{beta}
\end{equation}

Although the above mass and mixing angle intervals
are allowed by the ratios $R$,
one must make sure that the same allowed ranges do not
violate the experimental partial decay widths. This turns out to
be the case for the mass range $m_3 \gsim 215~\mbox{MeV}$, as will be shown in
Section IV--A.

\bigskip
In the following
we will examine
what would happen
to the above PPZ conclusions if 1) radiative corrections are taken
into account, 2) the constraint from $\overline{\Gamma}$'s (decay widths)
is imposed and 3) 1995 data are used
together with the $\overline{\Gamma}$ constraint and the radiative corrections.
Specifically, we will show that due to the accuracy of the
present data, it is important
to include
radiative corrections  and that the use of ratios $R$'s
alone without checking the partial decay widths
(i.e. $\overline{\Gamma}$'s)
could lead to overestimates of the allowed interval of the neutrino
parameters.

\subsection{Allowed Range for Mass}

In Fig.1 we have plotted the ranges for the values
of $m_3$ which are forbidden (denoted by solid lines)
by the ratios $R$'s,
by the $\overline{\Gamma}$'s and by the combination of
the two.
The heavy solid line represents the allowed region. In this plot,
it is assumed as in the case of the PPZ that
the other mass parameters are very small
($m_1\simeq m_2 \ll m_3$) and the mixing angles are varied over
the maximum interval $(0,\pi/2)$. The figure refers to three
cases:  1992 PDG data, without radiative corrections (RC), denoted by
92, 1992  PDG data with RC, denoted by 92RC
and 1995 data with RC, denoted by 95RC, respectively.
For each case we present the three
results, one with the $R$'s, one with the $\overline{\Gamma}$'s
and one with $R$'s and $\overline{\Gamma}$'s combined. For 92,
there is an allowed region for $m_3$ which does not include
$m_3=0$ when $R$'s alone are used. This is the PPZ result.
Most values of $m_3$ inside this
allowed region, however, violate the limits on the $\overline{\Gamma}$'s,
as can be seen in Fig.1.
It has to be emphasized that the range for $m_3$ allowed by the
$\overline{\Gamma}$'s alone does not include $m_3 = 0$, also.

If we combine the two results,
we find the following allowed range for $m_3$

\begin{equation}
178~\mbox{MeV} \lsim m_3 \lsim 215~\mbox{MeV}.
\label{naively}
\end{equation}

The inclusion of radiative corrections changes
the picture dramatically. As can be seen in Fig.1 (92RC),
the region allowed by $R$'s is considerably enlarged.
Also the $\overline{\Gamma}$ constraint is modified in an
important way, i.e. all the region of lower masses is now allowed,
including $m_3 = 0$.
When we combine the two constraints
($R + \overline{\Gamma}$) the PDG 92 data set is consistent
with the following finite (not including zero) mass range for $m_3$

\begin{equation}
140~\mbox{MeV} \lsim m_3 \lsim 210~\mbox{MeV}.
\label{m3:92}
\end{equation}
signaling new physics beyond the Standard Model.
It is, therefore, extremely interesting to repeat the
analysis with the 1995 data. It is to be noted that an agreement
of the above result with the PPZ's (Eq.(\ref{mass}))
is purely accidental.

As can be seen in 95RC in Fig 1, $R$'s alone
allow two different (disconnected) regions,
one of which includes $m_3 = 0$.
The inclusion of $\overline{\Gamma}$'s restricts these
intervals to the region of lower masses.
The allowed regions resulting from
the combined (both $R$'s and $\overline{\Gamma}$'s)
analyses based on the 1995 data with radiative corrections is
(for $m_1\simeq m_2 \ll m_3$)

\begin{equation}
m_3 \lsim 70~\mbox{MeV}
{}~~~~~\mbox{and}~~~~~
140~\mbox{MeV} \lsim m_3 \lsim 149~\mbox{MeV}.
\label{mass:lim}
\end{equation}

The above result is rather insensitive to the choice of
$m_1$ and $m_2$. For example, for $m_1 \lsim 20$~KeV and
$m_2 \lsim 1~\mbox{MeV}$, the above result remains unchanged.
Only if $m_2$ is of the order of a few MeV all the region between
$m_3 = 0$ and $m_3 \simeq 149~\mbox{MeV}$ is allowed.
Therefore, we conclude, based on the 1995 data, that
the 1992 PDG data were internally inconsistent
because of the poor data on $\tau$. Moreover, the accuracy of
the current data or even the 1992 PDG data warrants the inclusion
of the radiative corrections for any precision
analysis. Furthermore, one can see that the use of the $R$'s alone
without the constraint from $\overline{\Gamma}$'s can give rise to
overestimated allowed regions.

\subsection{Allowed Range for Mixing Angles }

So far we have reexamined and discussed the PPZ using the 1992 PDG data.
We have also carried out a similar analysis using
the 1995 data
with a conclusion that there
is an allowed window for $m_3$ including $m_3 = 0$, as well as
an isolated range of $m_3$ (for $m_1\simeq m_2 \ll m_3$).
In this Sub--Section we carry out a similar analysis for
mixing angles. The PPZ concludes, based on the ratios alone from
1992 PDG data, that
the angle $\beta$ has a finite allowed range,
$11.54^\circ \lsim \beta \lsim 12.82^\circ$,
whereas the angle $\gamma$
is restricted to $\gamma \lsim 4.05^\circ$, including zero.
This allowed region is shown in the $\beta$--$\gamma$ plane
as an area filled with circles in Fig.2a.
In addition, the constraint imposed by
the $\overline{\Gamma}$'s alone is indicated by the dotted region.
(It is to be pointed out that in  Fig.2 the mass parameters are
$m_1\simeq m_2 \simeq 0$ and $m_3=200~\mbox{MeV}$).
The region allowed by the two constraints combined is denoted by the
dark area.

Now, the inclusion of radiative corrections to the PPZ analysis
leads to changes in the allowed regions of Fig.2a.
The region allowed by $R$'s alone is enlarged
and moved towards the origin, but still it does not include the
origin that corresponds to the case of
neutrinos without mixing. This is shown by circles in Fig.2b. Also
shown in Fig.2b is the allowed region based on $\overline{\Gamma}$'s
alone, with the radiative corrections included (dots). The darker area
is the region allowed by the $R$'s and $\overline\Gamma$'s combined.
Therefore, the common allowed region is

\begin{eqnarray}
& & 4^\circ \lsim \beta \lsim 10^\circ  \\
& & \gamma \lsim 7.2^\circ
\end{eqnarray}

When we use the 1995 data (with radiative corrections), the situation
again changes dramatically. Fig.2c shows the allowed region
in the $\beta$--$\gamma$ plane which is
obtained by using the ratios $R$ alone
(circles). The allowed region is shown to move
farther away from the origin.
The addition of the constraint coming from the
$\overline{\Gamma}$'s completely washes out
the region, i.e. no allowed region.
This is self--evident because the mass $m_3=200$ MeV is not
allowed by the 1995 data, as can be seen in Eq.(\ref{mass:lim}).
Thus, even in the case of the 1995 data, neglecting the
$\overline\Gamma$ constraint could lead to erroneous conclusions.

\section{Analysis and Results}

In this Section we present the results of
a more detailed analysis of the
limits that can be set on the neutrino masses and mixing angles
by using the 1995 data and with the inclusion of radiative
corrections. It is to be pointed out here
that a complete, combined analysis of the masses ($m_1$, $m_2$
and $m_3$) and mixing angles ($\theta$, $\beta$ and $\gamma$)
is very much involved and is beyond
the scope of this paper. Even the presentation
of results of such an analysis
would be problematic.
Therefore, we have  simplified  the analysis by fixing
some parameters and varying others.
In order to see various correlations among the masses and the mixing
angles, we present several allowed regions in the two dimensional
plots for several combinations of mass and mixing angle.

First, we present the absolute upper limits on
the three neutrino masses, $m_1$, $m_2$, and $m_3$, independent of
the values of mixing angles, $\theta$, $\beta$ and
$\gamma$. We have obtained  these
limits by varying the mixing angles over the entire interval
between  0 and $\pi/2$
and by taking into account all the constraint which
we have discussed in the previous Section ($R$'s and
$\overline{\Gamma}$'s) and the radiative corrections.
(It should be stated that we are \underline{not} carrying out
a statistical analysis of all the relevant data.)
Instead of three dimensional plots, we present, in Fig.3,
the allowed region (dotted area) in the $m_1$--$m_2$
plane  and  the allowed region  in $m_2$--$m_3$ plot in  Fig.4.
{}From these plots, we can set the
following {\em absolute} upper limits on the neutrino masses
(based on $\overline\Gamma$ and $R$ constraint, at 1--$\sigma$
level)

\begin{equation}
\begin{array}{l}
m_1 \lsim 100 ~\mbox{KeV} ,    \\
m_2 \lsim 7.5 ~\mbox{MeV} ,    \\
m_3 \lsim 149    ~\mbox{MeV}.
\label{mass:method1}
\end{array}
\end{equation}
where the limits on $m_1$ and $m_2$ are mainly due to the $R$'s, whereas
the limit on $m_3$ comes from the $\overline\Gamma$ constraint.
The limit on $m_3$ has already been mentioned in the previous Section.
Although the limit on $m_1$ is rather poor,
the limit on $m_2$ is
larger by a factor of less than thirty than the latest limit from the
kinematical analysis of the $ \pi \rightarrow \mu + \nu_\mu$ decay.
Similarly, the limit on
$m_3$ is larger only by a factor of six. It is quite interesting that
the accuracy of the present data on the decay rates and branching
ratios is already sufficiently good enough to set limits on
$m_2$ and $m_3$ which
agree, in one order of magnitude, with the results
{}from more involved kinematical determination.
The important difference between the upper limits given in
Eq.(\ref{mass:method1}) and those in Eq.(\ref{kin}) is that the
former is valid independently of mixing angles whereas the latter
is valid only for the case of no mixing.
Furthermore, the improvement in the data from 1992 is
obvious from the conclusion that the upper limit on $m_3$ is set to 149 MeV
and the limits are consistent with massless neutrinos with no mixing,
implying the internal consistency of the data.

Next, we discuss some correlations among the masses and the mixing
angles.
The first example to be presented is
the allowed region in the $m_2$--$\sin^2{\theta}$ plot shown in Fig.5.
In this plot, we have set $m_3$ to be 24 MeV and
$\nu_3$ is assumed to be decoupled, due to its heavy mass, from
$\nu_1$ and $\nu_2$ so that $\beta = \gamma = 0$. Also, for definiteness,
we have taken $m_1 = 5$ eV, but the conclusion remains unchanged
as long as $m_1$ is less than $\sim 20 \mbox{KeV}$.
The solid and dashed
lines delimit allowed regions based on  the use of
$\overline{\Gamma}$'s and the  constraint from $R$'s, respectively.
The allowed area is denoted by dots. Fig.5 shows that
the low--angle regime is constrained mainly by
$R^{\mu e}_{\tau e}$ whereas
$R^{\pi e}_{\pi \mu}$ is more effective in limiting
the large--angle area (say $\sin^2\theta \gsim 10^{-4}$).
The allowed region shown in Fig.5 is insensitive to
values of
$m_1$ and $m_3$, as long as they are $m_1 \lsim 20 ~\mbox{KeV}$
and $m_3 \lsim 50~\mbox{MeV}$.
The so--called small-- and large-- angle solutions of the solar neutrino
deficit
based on the MSW effect, which requires $m^2_2 - m^2_1 \simeq
6 \times 10^{-6} \mbox{eV}^2$ and $\sin^2(2\theta) \simeq 7 \times 10^{-3}$,
and $m^2_2 - m^2_1 \simeq 8 \times 10^{-6} \mbox{eV}^2$ and
$\sin^2(2\theta) \simeq 0.6$, respectively,
are well within the allowed region in this plot.

In the next example, we assume that
$\nu_1$ is too light to couple with $\nu_2$ and $\nu_3$. That is,
only $\nu_2$ and $\nu_3$ are mixed with angle $\gamma$. We have also set
$m_1 \ll m_2 = 270~\mbox{KeV}$. Fig.6 shows the allowed region
in the  $m_3$--$\sin^2\gamma$ plane.
Here, the small-- and large-- angle regions are constrained by
$R^{\mu e}_{\tau e}$, whereas the intermediate region
($\sin^2\gamma \sim 10^{-4}$--$10^{-2}$) is constrained by
$R^{\pi e}_{\pi \mu}$.
The neutrino oscillation solution of the
atmospheric neutrino anomaly as observed
by Kamiokande and others,
which favors $m^2_3 - m^2_2 \simeq 10^{-2} \mbox{eV}^2$
and $\sin^2(2\theta) \simeq 1$, is well within the allowed
region in Fig.6. The size and the shape of the
allowed region are insensitive to
the assumed values of $m_1$ and $m_2$ as long as $m_1 \lsim 20 ~\mbox{KeV}$
and $m_2 \lsim 1 ~\mbox{MeV}$.

Although the case of a $\nu_1$-- $\nu_3$ mixing
is unnatural in the framework of natural hierarchy
of neutrino masses, we present it in Fig.7.
In this Figure the allowed region in the
$m_3$ -- $\sin^2 \beta$ plane is shown. Here, as in the previous case
we have set $m_1 \ll m_2 = 270~\mbox{KeV}$.
In this case, the most restricted
limit is always imposed by $R^{\pi e}_{\pi \mu}$. Again,
the allowed region
does not change significantly as long as $m_1 \lsim 20 ~\mbox{KeV}$
and $m_2 \lsim 1 ~\mbox{MeV}$.

\bigskip

In the previous figures Figs.5--7 the limits were
obtained in the special cases in which two of the three mixing
angles were kept fixed at zero, i.e. only one pair of
neutrinos is mixed. To show how sensitive
these limits are to the fixed angles and correlations
among the limits, we present
 in Figs.8--10 the cross sections (with one angle fixed)
of the three--angle parameter space for fixed values of the
masses, i.e. $\sin^2 \gamma$--$\ \sin^2\beta$
in Fig.8, $\sin^2 \gamma$--$ \sin^2\theta$
in Fig.9 and $\sin^2 \beta$--$\ \sin^2\theta$
in Fig.10, respectively. In each figure, the values of the masses
are fixed as $m_1 \ll m_2=270 ~\mbox{KeV}$ and
$m_3=24~\mbox{MeV}$. It is interesting to note
that for the above set of the masses,
the most severely constrained angle is $\beta$,
whereas the least constrained is  $\gamma$. The limits
are:

\begin{equation}
\begin{array}{l}
\sin^2 \theta \lsim 3.6 \cdot 10^{-3}  \\
\sin^2 \beta  \lsim 4.6 \cdot 10^{-7}  \\
\sin^2 \gamma \lsim 7.0 \cdot 10^{-3}.
\end{array}
\label{angle:1}
\end{equation}

As we decrease the values of the masses used, the allowed region
in each figure increases, eventually covering the entire space.
Hence, no meaningful limit can be obtained, as expected.
In order to demonstrate this sensitivity, we have shown in
Figs.8--10 the extended allowed regions (bounded
by the dashed lines) for the
following values of the masses: $m_1 \ll m_2=10 ~\mbox{KeV}$ and
$m_3=1~\mbox{MeV}$.
In this case, the limits on the mixing angles
substantially increase to

\begin{equation}
\begin{array}{l}
\sin^2 \theta \lsim 1  \\
\sin^2 \beta  \lsim 2.6 \cdot 10^{-4}  \\
\sin^2 \gamma \lsim 1.
\end{array}
\label{angle:2}
\end{equation}

\section{Summary and Conclusions}

We have analyzed  and compared the 1992 and 1995 data on the
$\pi$, $\mu$ and $\tau$ decays in the framework of
three generations of massive neutrinos with associated mixing.
First we have confirmed the surprising result of PPZ based on
the 1992 PDG data that when only the ratios, $R$'s,  are
used without radiative corrections, the 1992 data are inconsistent
with the picture of massless neutrinos with no mixing, signaling
new physics beyond the Standard Model!
More specifically, PPZ has shown
that when $m_1$ and $m_2$ are assumed to be much less than
$m_3$, $m_3$ is found to be  within the interval
$155~\mbox{MeV} \lsim m_3 \lsim 225~\mbox{MeV}$ and the
$\nu_1$ -- $\nu_3$ mixing angle, $\beta$, in the interval,
$11.54^\circ \lsim \beta \lsim 12.84^\circ$.
This isolated allowed region survives even if we introduce the
$\overline{\Gamma}$ constraint.

Again, the 1992 PDG data set with radiative corrections reproduces
the results that agree qualitatively with those of PPZ. The allowed
range of $m_3$ is $140~\mbox{MeV} \lsim m_3 \lsim 210~\mbox{MeV}$
and the allowed mixing angles are
$4^\circ \lsim \beta \lsim 10^\circ$ and $\gamma \lsim 7.2^\circ$.
This clearly shows
that the 1992 PDG data set suggests massive and mixed neutrino.

In order to see if this rather surprising result still remains valid
or not with the improved data of 1995, we have carried out a
comprehensive analysis of the 1995 data
by using both the ratios, $R$'s, and decay widths, $\overline{
\Gamma}$'s, and by including the radiative corrections. The 1995 data
are shown to be  internally consistent, and furthermore
consistent with the picture of massless neutrinos with no mixing.

Limits on the masses derived from the analysis are as follows:

\begin{equation}
\begin{array}{l}
m_1 \lsim 100 ~\mbox{KeV} ,    \\
m_2 \lsim 7.5 ~\mbox{MeV} ,    \\
m_3 \lsim 149    ~\mbox{MeV}.
\label{absolute:lim}
\end{array}
\end{equation}

These bounds on the masses are such that the imposed constraint
($R$'s and $\overline\Gamma$'s) are fulfilled in their 1--$\sigma$
intervals.

Although the above limits are less stringent than those in Eq.(\ref{kin})
{}from kinematical determinations, it is important to note that the
limits given in Eq.(\ref{absolute:lim}) are completely independent of
mixing angles. Therefore, if neutrinos are massive and mixed, their masses
can be heavier than the limits given in Eq.(\ref{kin}). Of course,
such heavy neutrinos, if stable, are not allowed by the well--known
cosmological limits, $\sum_i m(\nu_i) \lsim 20 \sim 30~\mbox{eV}$
(with $h=(H_0 \mbox{~s~Mpc})/\mbox{100 Km}\simeq 0.5$,
where $H_0$ is the Hubble constant).
If unstable, the decay of neutrinos must be
such that it should not disturb the standard nucleosynthesis scenario
and should not violate the observed limits on cosmic electromagnetic
wave backgrounds.

It is not possible to obtain absolute limits on the mixing angles,
because they strongly depend on the input values of the masses.
Examples of limits on the angles for definite values of the masses
have been derived and reported in the previous Section.

To conclude, it is gratifying that the accuracy of the current data
is already good enough to set limits on
$m_2$ and $m_3$ which agree, in the order of magnitude,
with the results from more involved kinematical determination,
although the limits on the mixing angles are still rather poor.
Further improvements of the data on $\tau$, $m_W$ and
$m_Z$ in the future may significantly improve the limits.

\acknowledgments
The authors would like to thank B. Blumenfeld, J. Bagger,
A. Falk, P. Gambino, L. Madansky
and G. Passarino for helpful discussions. The correspondence with
O.L.G. Peres, V. Pleitez, R. Zukanovich Funchal and R.E. Shrock
was stimulating and very helpful. NF wishes to express his
gratitude to the Fondazione ``A. Della Riccia'' for a fellowship
and CWK wishes to thank
the Dipartimento di Fisica Teorica, Universit\`a di Torino
and INFN -- Sezione di Torino for the hospitality extended to him
where this work began.
This work was supported in part by the
National Science Foundation, USA and by the Ministero della Ricerca
Scientifica e Tecnologica, Italy.

\appendix
\section{Formulas for pion and lepton decays}

In this Appendix
we present, for completeness, the expressions for the pion
and lepton decays for the general case of three massive
neutrinos with mixing. Radiative corrections to these processes are
included with a brief comment.

\subsection{\bf Pion decay}

The partial decay rate of the pion into two leptons is given by
\begin{equation}
\Gamma(\pi \longrightarrow l \bar\nu_l) =
\frac{G^2 f_\pi^2 U_{ud}^2 m_\pi^3}{8 \pi}
\cdot {\cal R}_{\pi l} \cdot
\sum_{i=1}^3 |V_{li}|^2 P_i^{\pi l},
\label{pion:g}
\end{equation}
where the phase space--matrix element factor, $ P_i^{\pi l}$, is given by
\begin{equation}
P_i^{\pi l} =
\theta(m_\pi-m_l-m_i)
\cdot
[\delta_{l\pi}^2 + \delta_{i\pi}^2 -
(\delta_{l\pi}^2 - \delta_{i\pi}^2)^2]
\lambda^{1/2}(1,\delta_{l\pi}^2,\delta_{i\pi}^2).
\end{equation}
In the above,
$G$ is the Fermi constant, $f_\pi$ is the pion decay constant,
$U$ is the mixing matrix in the quark sector,

\begin{eqnarray}
\delta_{l\pi} &=& \frac{m_l}{m_\pi}               \\
\delta_{i\pi} &=& \frac{m_i}{m_\pi} ~~ (i=1,2,3)
\end{eqnarray}
and $\lambda$ is the standard kinematical function

\begin{equation}
\lambda(x,y,z) = x^2 + y^2 + z^2 - 2 (xy+yz+xz).
\end{equation}

The quantity ${\cal R}_{\pi l}$ in Eq.(\ref{pion:g})
describes the {\em leading} radiative corrections
to the pion decay process \cite{PDG94,sirlin}
given by

\begin{eqnarray}
{\cal R}_{\pi l} &=&
\left[ 1 + \frac{2 \alpha}{\pi}
\ln\left(\frac{m_Z}{m_\rho}\right)\right]
\left[1+\frac{\alpha}{\pi} F(\delta _{l \pi}) \right]
\nonumber       \\
& & \times \left\{ 1 - \frac{\alpha}{\pi}
\left[\frac{3}{2}\ln\left(\frac{m_\rho}{m_\pi}\right) +
C_1 +
C_2 \frac{m_l^2}{m_\rho^2} \ln\left(\frac{m_\rho^2}{m_l^2}\right) +
C_3 \frac{m_l^2}{m_\rho^2} + \dots
\right]\right\}
\label{pi:rc}
\end{eqnarray}
where

\begin{eqnarray}
F(x) &=&
3\ln x + \frac{13-19 x^2}{8(1-x^2)} -
\frac{8-5x^2}{2(1-x^2)^2}x^2\ln x
\nonumber \\
& &
\mbox{} - 2 \left( \frac{1+x^2}{1-x^2}\ln x +1 \right)\ln(1-x^2) +
2\left( \frac{1+x^2}{1-x^2} \right)L(1-x^2)
\end{eqnarray}
Here, $m_\rho = 796$~MeV is
the $\rho$ meson mass, $m_Z$  the $Z$ boson mass and $\alpha$ is the fine
structure constant. Also, in the above,
$L(z)$ is defined by

\begin{equation}
L(z) = \int_0^z \frac{\ln(1-t)}{t} dt.
\end{equation}

The first square bracket in Eq. (\ref{pi:rc}) represents the
electromagnetic short--distance correction. Its value is
slightly modified when higher--order effects and QCD corrections
are taken into account, i.e. \cite{sirlin}

\begin{equation}
\left[ 1 + \frac{2 \alpha}{\pi}
\ln\left(\frac{m_Z}{m_\rho}\right)\right] \longrightarrow 1.0232.
\end{equation}
The second and the third brackets denote the QED corrections to
the decay of a pointlike pion \cite{sirlin,kinoshita}. Following
a general practice, we neglect
the terms with  the $C_i$'s, whose numerical values have large
uncertainties \cite{sirlin}. Therefore, we use the following
simplified expression

\begin{equation}
{\cal R}_{\pi l} = 1.0232 \cdot
\left[1+\frac{\alpha}{\pi} F(\delta _{l \pi}) \right] \cdot
\left[ 1 - \frac{3\alpha}{2\pi}
\ln\left(\frac{m_\rho}{m_\pi}\right)\right].
\end{equation}

Note that ${\cal R}_{\pi l}$ depends both on the pion and the
lepton mass.

\subsection{\bf Lepton decay}

The partial decay rate of a lepton into three leptons is given by
\begin{equation}
\Gamma(l' \longrightarrow l \bar \nu_{l} \nu_{l'}) =
\frac{G^2 m_{l'}^5}{192 \pi^3}\cdot {\cal R}_{l'} \cdot
\sum_{i,j=1}^3 |V_{l'i}|^2 |V_{lj}|^2 P_{ij}^{l'l},
\end{equation}
where

\begin{equation}
P_{ij}^{l'l} = \theta(m_{l'}-m_l-m_i-m_j) \cdot
2 \int_{x_{\mbox{\scriptsize min}}}^{x_{\mbox{\scriptsize max}}}
(x^2 - 4\delta_{ll'}^2)^{1/2} {\cal M} dx
\end{equation}
and

\begin{eqnarray}
{\cal M} &=&
{\cal A}^{1/2}[x(1+\delta_{ll'}^2-x){\cal A}
+ (2-x)(x-2\delta_{ll'}^2){\cal B}],
\\
{\cal A} &=&
\frac{1}{(1+\delta_{ll'}^2-x)^2} \nonumber
\\
& & \times
[(1+\delta_{ll'}^2-x)^2-
2(1+\delta_{ll'}^2-x)(\delta_{il'}^2+\delta_{jl'}^2) +
(\delta_{il'}^2-\delta_{jl'}^2)^2],
\\
{\cal B} &=&
\frac{1}{(1+\delta_{ll'}^2-x)^2} \nonumber
\\
& & \times
[(1+\delta_{ll'}^2-x)^2+
(1+\delta_{ll'}^2-x)(\delta_{il'}^2+\delta_{jl'}^2) -
2(\delta_{il'}^2-\delta_{jl'}^2)^2].
\end{eqnarray}

In the above

\begin{eqnarray}
\delta_{ll'}     &=& \frac{m_l}{m_{l'}} \\
\delta_{il'}     &=& \frac{m_i}{m_{l'}} ~~~(i=1,2,3)\\
x_{\mbox{\scriptsize min}} &=&
2\delta_{ll'}       \\
x_{\mbox{\scriptsize max}} &=&
1+\delta_{ll'}^2-(\delta_{il'}+\delta_{jl'})^2.
\end{eqnarray}

The quantity ${\cal R}_{l'}$ describes the {\em leading}
radiative corrections to the lepton decay process \cite{ldcy}
which are given by

\begin{equation}
{\cal R}_{l'} =
\left [
1 + \frac{\alpha}{2\pi}\left( \frac{25}{4}-\pi^2 \right)
\right ]
\left (
1 + \frac{3}{5}\frac{m_{l'}^2}{m_W^2}
\right ),
\end{equation}
where $m_W$ is the $W$ boson mass.
Note that ${\cal R}_{l'}$ depends only on the mass of the
decaying lepton, but not on the mass of decay products.

\figure{FIG.1~ Allowed (thick solid lines) and forbidden (solid lines)
intervals of $m_3$ by $R$'s, $\overline{\Gamma}$'s
and $R$'s and $\overline{\Gamma}$'s
combined, respectively, for 1992 and 1995 data sets
($m_1\simeq m_2 \ll m_3$ is assumed).
Mixing angles are varied over the interval $(0,\pi/2)$.
Label 92 refers to the result of 1992 PDG data,
without radiative corrections in the calculation; 92RC and 95RC denote
the results of 1992 PDG and 1995 data sets with
radiative corrections, respectively. The mass scale is in arbitrary units.
\label{figure1}}

\figure{FIG.2~ Allowed (dotted) regions in the
$\beta$--$\gamma$ plane
for 1992 and 1995 data sets with $m_1\simeq m_2\simeq 0$
and $m_3=200~\mbox{MeV}$:
(a) allowed region by $R$'s alone (circles),
$\overline{\Gamma}$'s alone (dots) and $R + \overline\Gamma$
combined (dark area) of 1992 PDG data, (b)
allowed region by $R$'s alone (circles) and
by $\overline{\Gamma}$'s alone (dots) of 1992 PDG data with
radiative corrections (the darker area is the region allowed
both by the $\overline{\Gamma}$'s and the $R$'s),
(c) allowed region by $R$'s alone (circles) of 1995
data with radiative corrections.
\label{figure2}}

\figure{FIG.3~ Allowed (dotted) region in the $m_1$--$m_2$ plane
for 1995 data with radiative corrections. Both $R$'s and
$\overline{\Gamma}$'s constraints are imposed.
Mixing angles are varied over the interval $(0,\pi/2)$.
\label{figure3}}

\figure{FIG.4~ Allowed (dotted) region in the $m_2$--$m_3$ plane
for 1995 data with radiative corrections. Both $R$'s and
$\overline{\Gamma}$'s constraints are imposed.
Mixing angles are varied over the interval $(0,\pi/2)$.
\label{figure4}}

\figure{FIG.5~ Allowed (dotted) region in the $m_2$--$\sin^2\theta$ plane
for 1995 data with radiative corrections and
$m_1=5~\mbox{eV}$, $m_3=24~\mbox{MeV}$ and $\beta=\gamma=0$.
The regions below the solid and the dashed lines
are allowed by $\overline{\Gamma}$'s alone and $R$'s alone,
respectively.
\label{figure5}}

\figure{FIG.6~ Allowed (dotted) region in the $m_3$--$\sin^2\gamma$ plane
for  1995 data with radiative corrections and
$m_1=5~\mbox{eV}$, $m_2=270~\mbox{KeV}$ and $\theta=\beta=0$.
The regions below the solid and the dashed lines
are allowed by $\overline{\Gamma}$'s alone and $R$'s alone,
respectively.
\label{figure6}}

\figure{FIG.7~ Allowed (dotted) region in the $m_3$--$\sin^2\beta$ plane
for 1995 data with radiative corrections, and
$m_1=5~\mbox{eV}$, $m_2=270~\mbox{KeV}$ and $\theta=\gamma=0$.
The regions below the solid and the dashed lines
are allowed by $\overline{\Gamma}$'s alone and $R$'s alone,
respectively.
\label{figure7}}

\figure{FIG.8~ Allowed regions in the $\sin^2\gamma$--$\sin^2\beta$
plane for 1995 data with radiative corrections.
The dotted area inside the solid line is for
$m_1\ll m_2=270~\mbox{KeV}$ and $m_3=24~\mbox{MeV}$.
The area on the left of the dashed line is for
$m_1\ll m_2=10~\mbox{KeV}$ and $m_3=1~\mbox{MeV}$.
Note the enlargement of the allowed region as
masses decrease.
\label{figure8}}

\figure{FIG.9~ Allowed regions in the $\sin^2\gamma$--$\sin^2\theta$
plane for 1995 data with radiative corrections.
Notation and interpretation are the same as in Fig.8.
\label{figure9}}

\figure{FIG.10~ Allowed regions in the $\sin^2\beta$--$\sin^2\theta$
plane for 1995 data with radiative corrections.
Notation and interpretation are the same as in Fig.8.
\label{figure10}}

\renewcommand{\arraystretch}{1.5}
\begin{table}
%\centering
\caption{List of 1992 and 1995 data sets.
The entries with asterisks denote data which have been improved
by a significant amount.}
%\vspace{.35truecm}
\begin{tabular}{|c|c|cc|cc|}
\mbox{ } & \mbox{ } &
{\bf 1992 data} & \mbox{ } &
{\bf 1995 data} & \mbox{ } \\
\tableline
\mbox{ } &
$m_e$ &
$(0.51099906 \pm 0.00000015) \mbox{MeV} $ & \mbox{ } &
\mbox{same as 1992} & \mbox{ } \\
\mbox{ } &
$m_\mu$ &
$(105.658389 \pm 0.000034) \mbox{MeV} $  & \mbox{ } &
\mbox{same as 1992}  & \mbox{ } \\
\mbox{*} &
$m_\tau$ &
$(1784.1 \pm 3.6) \mbox{MeV} $  & \mbox{ } &
$(1776.96 \pm 0.31) \mbox{MeV} $  & \mbox{ } \\
\mbox{ } &
$m_\pi$ &
$(139.5679 \pm 0.0007) \mbox{MeV}  $ & \mbox{ } &
$(139.56995 \pm 0.00035) \mbox{MeV}   $ & \mbox{ } \\
\mbox{ } &
$\tau_\mu$ &
$(2.19703 \pm 0.00004) \cdot 10^{-6} ~\mbox{s} $ & \mbox{ } &
\mbox{same as 1992} & \mbox{ } \\
\mbox{*} &
$\tau_\tau$ &
$(305 \pm 6) \cdot 10^{-15} ~\mbox{s}     $   & \mbox{ } &
$(291.6 \pm 1.5) \cdot 10^{-15} ~\mbox{s} $   & \mbox{ } \\
\mbox{*} &
$m_Z$  &
$(91.173 \pm 0.020) \cdot 10^{3} \mbox{MeV} $  & \mbox{ } &
$(91.1884 \pm 0.0022) \cdot 10^{3} \mbox{MeV} $ & \mbox{ } \\
\mbox{*} &
$m_W$  &
$(80.22 \pm 0.26) \cdot 10^{3} \mbox{MeV} $  & \mbox{ } &
$(80.410 \pm 0.180) \cdot 10^{3} \mbox{MeV} $ & \mbox{ } \\
\mbox{ } &
$G_\mu$  &
$(1.16639 \pm 0.00002) \cdot 10^{-11} \mbox{MeV}^{-2}  $  & \mbox{ } &
\mbox{same as 1992}  & \mbox{ } \\
\mbox{*} &
$G$  &
$(1.162 \pm 0.029) \cdot 10^{-11} \mbox{MeV}^{-2}  $  & \mbox{ } &
$(1.174 \pm 0.022) \cdot 10^{-11} \mbox{MeV}^{-2}  $  & \mbox{ }  \\
\mbox{*} &
$\mbox{BR}(\pi \longrightarrow e \bar\nu_e)$  &
$(1.218 \pm 0.014) \cdot 10^{-4}$               & \mbox{ } &
$(1.230 \pm 0.004) \cdot 10^{-4}$               & \mbox{ } \\
\mbox{ } &
$\mbox{BR}(\pi \longrightarrow \mu \bar\nu_\mu)$  &
$0.9998782 \pm 0.0000014$                           & \mbox{ } &
$0.9998770 \pm 0.0000004$                           & \mbox{ } \\
\mbox{ } &
$\mbox{BR}(\mu \longrightarrow e \bar\nu_e \nu_\mu)$ &
$1$                                                    & \mbox{ } &
\mbox{same as 1992} & \mbox{ } \\
\mbox{*} &
$\mbox{BR}(\tau \longrightarrow e \bar\nu_e \nu_\tau)$ &
$0.1793 \pm 0.0026$                                      & \mbox{ } &
$0.1779 \pm 0.0009$                                      & \mbox{ } \\
\mbox{*} &
$\mbox{BR}(\tau \longrightarrow \mu \bar\nu_\mu \nu_\tau)$  &
$0.1758 \pm 0.0027$                                           & \mbox{ } &
$0.1733 \pm 0.0009$                                           & \mbox{ } \\
\tableline
\mbox{*} &
$\overline{\Gamma}^{\mu e}$  &
$1.003 \pm 0.050             $ & \mbox{ } &
$0.983 \pm 0.037             $ & \mbox{ } \\
\mbox{*} &
$\overline{\Gamma}^{\tau e}$  &
$0.944 \pm 0.053                  $ & \mbox{ } &
$0.979 \pm 0.037                  $ & \mbox{ } \\
\mbox{*} &
$\overline{\Gamma}^{\tau \mu}$  &
$0.925 \pm 0.052                $ & \mbox{ } &
$0.954 \pm 0.036                $ & \mbox{ } \\
\mbox{*} &
$R^{\pi e}_{\pi \mu}$  &
$(1.218 \pm 0.013) \cdot 10^{-4}  $ & \mbox{ } &
$(1.230 \pm 0.004) \cdot 10^{-4}  $ & \mbox{ } \\
\mbox{*} &
$R^{\mu e}_{\tau e}$  &
$1.063  \pm 0.028                   $ & \mbox{ } &
$1.0038 \pm 0.0073                  $ & \mbox{ } \\
\mbox{*} &
$R^{\tau e}_{\tau \mu}$  &
$1.020  \pm 0.022                $ & \mbox{ } &
$1.0265 \pm 0.0074                $ & \mbox{ } \\
\end{tabular}
\label{tab:data}
\end{table}

\end{document}